Physics and chemistry review of layered chalcogenide superconductors

Keita Deguchi[1, 2], Yoshihiko Takano[1, 2], and Yoshikazu Mizuguchi[1, 3, *]

1. National Institute for Materials Science, 1-2-1, Sengen, Tsukuba, 305-0047, Japan.

2. Graduate School of Pure and Applied Sciences, University of Tsukuba, 1-1-1 Tennodai, Tsukuba, 305-8571, Japan

3. Department of Electrical and Electronic Engineering, Tokyo Metropolitan University, 1-1, Minami-osawa, Hachioji, 192-0397, Japan.

* Corresponding author: Yoshikazu Mizuguchi (e-mail: mizugu@tmu.ac.jp)

Abstract

Structural and physical properties of layered chalcogenide superconductors are summarized. In particular, we review the remarkable properties of the Fe-chalcogenide superconductors, FeSe and FeTe-based materials. Furthermore, we introduce the recently-discovered new $BiS_2$-based layered superconductors and discuss its prospects.

## 1. Introduction to layered chalcogenide superconductors

Layered materials have provided us with many interesting fields in physics and chemistry. Owing to their two-dimensional crystal structure and electronic states, anomalous electronic and magnetic properties have often been observed. In particular, exotic superconductivity is likely to prefer such a layered crystal structure. For example, superconductivity at a high transition temperature ($T_c$) has been achieved in layered materials, such as cuprates [1-4], Fe-based [5-13], and $MgB_2$ [14,15] superconductors. With the respect to this fact, we expect that a new high-$T_c$ superconducting family will be found in novel layered materials.

Among the layered superconductors, the "chalcogenides" are one of the notable groups, because of the variety of materials and the observation of exotic superconductivity. Sulfur (S), selenium (Se) and tellurium (Te) are categorized as chalcogens. Here we introduce the structural and physical properties of some layered chalcogenide superconductors. $CdI_2$-type $TiSe_2$ has a simple layered structure composed of a stacking of $TiSe_2$ layers as depicted in Fig. 1(a), and exhibits a charge-density-wave (CDW) state. When transition metals are intercalated into the interlayer sites or external pressures are applied, the CDW state is suppressed and superconductivity is induced [16,17]. A topological insulator $Bi_2Se_3$ also possesses a layered structure with a Van-der-Waals gap as displayed in Fig. 1(b). As the same in the case of $TiSe_2$, $Bi_2Se_3$ becomes superconducting by Cu intercalations in the interlayer sites [18,19]. One of the notable characteristics of chalcogenides is crystallization of a simple layered structure with Van-der-Waals gaps. In such a structure, ions can be easily intercalated to the interlayer sites and dramatically change the physical properties of the chalcogenide layers.

The most remarkable layered chalcogenides are the Fe chalcogenides, FeSe and FeTe, which are the simplest Fe-based superconductors. The crystal structure of FeSe is shown in Fig. 1(c). FeSe exhibits a superconducting transition around 10 K, and shows a dramatic increase of $T_c$ up to 37 K under high pressure. In contrast to FeSe, FeTe undergoes an antiferromagnetic transition at 70 K. However, a partial substitution of Te by S or Se suppresses the antiferromagnetic ordering and induces superconductivity. This family is very interesting because the physical properties dramatically change with changing local crystal structure upon the co-valent substitutions of S, Se and Te. Recently, superconductivity above 40 K was observed in metal- or molecule-intercalated FeSe. To that respect, studies on not only fundamental physics but also applications using Fe-chalcogenide superconductors will be actively addressed. In the second section, we summarize the physical and structural properties of the Fe-chalcogenide superconductors.

Very recently, we discovered a new superconducting family of the layered bismuth sulfides [20-22]. In this family, the $BiS_2$ layers are the common superconducting structure essential for superconductivity as the $CuO_2$ planes in the cuprates and the $Fe_2An_2$ (An: Anions of P, As, S, Se or Te) layers in the Fe-based family. In the third section, we introduce the discovery of the novel $BiS_2$-based layered superconducting family and the latest results.

## 2. Fe chalcogenides

In 2008, Kamihara *et al*. reported superconductivity in Fe-based compound $LaFeAsO_{1-x}F_x$ [5]. Although the parent compound LaFeAsO is an antiferromagnetic metal, the F substitution suppresses the magnetic ordering and induces superconductivity with a $T_c$ as high as 26 K. As shown in Fig. 2(a), LaFeAsO has a layered structure with a stacking of the blocking $La_2O_2$ layers and the superconducting $Fe_2As_2$ layers. Many FeAs-based compounds analogous to LaFeAsO, for example $SmFeAsO_{1-x}F_x$ and $Ba_{1-x}K_xFe_2As_2$, were found to be superconducting with a maximum $T_c$ as high as 55 K [7,8]. All the Fe-based superconductors share a common layered structure based on the planar layer of an Fe square lattice. In Fe-pnictide superconductors, blocking layers with alkali, alkali-earth or rare-earth and oxygen/fluorine are alternatively stacked with Fe-As conduction layers [5-9, 23-26].

Superconductivity in layered Fe chalcogenides was initially found in FeSe by Hsu *et al*. soon after the discovery of superconductivity in the LaFeAsO system [10]. Contrary to the FeAs-based superconductors, FeSe is composed of only superconducting $Fe_2Se_2$ layers as shown in Fig. 2(b). The absence of a blocking layer leads Fe-chalcogenides to be the simplest crystal structure among the Fe-based superconductors. Furthermore, $Fe_{1+d}Te$, which possesses a crystal structure analogous to FeSe as depicted in Fig. 2(c), exhibits antiferromagnetic ordering as observed in the parent (non-doped) phases of FeAs-based superconductors. Thus, the Fe-chalcogenides are the key materials for elucidating the mechanism of Fe-based superconductivity [10-12].

## 2-1. FeSe

PbO-type FeSe is well known as a commercial material, but the discovery of superconductivity in this compound was triggered by Fe-based superconductors reported in 2008 [10]. PbO-type FeSe is composed of only $Fe_2Se_2$ layers with a tetragonal structure (space group: *P4/nmm*). It has an iron-square planar sublattice equivalent to that of the iron pnictides and its crystal structure is the simplest among the Fe-based superconductors. The PbO structure is one of the stable phases of the Fe-Se binary compounds [27]. One can therefore obtain a polycrystalline sample of PbO-type FeSe using a conventional solid state reaction method. However, FeSe synthesized by the solid-state reaction at high temperatures partially contains the NiAs-type (hexagonal) FeSe phase. To obtain a single phase of tetragonal FeSe, low temperature annealing around 300-400°C, which transforms the hexagonal to the tetragonal phase, is required [28]. FeSe exhibits a structural transition from tetragonal to orthorhombic phase (*Cmma*) at 70-90 K. This structural transition is not accompanied by any magnetic transition and superconductivity occurs in the orthorhombic phase below ~ 12 K [10, 28-31]. Figure 3 displays the temperature dependence of resistivity and the insets show the resistivity under the magnetic fields and the estimated upper critical field ($\mu_0 H_{c2}$) [10]. The resistivity became zero below 8K, and the high $\mu_0 H_{c2}$, the common features of Fe-based superconductivity, were observed. Although the $T_c$ of FeSe at ambient pressure is relatively low as compared to the other Fe-based superconductors, FeSe shows a dramatic enhancement of $T_c$ under high pressures [32-35].

Fig. 4(a) and (b) show the temperature dependence of resistivity for FeSe under high pressure measured using a piston cylinder cell and an indenter cell, respectively [32, 34]. The pressure dependence of $T_c$ of FeSe shows an anomalous behavior. With increasing pressure, the pressure dependence of $T_c$ shows an anomaly around 2 GPa, and then exhibits a steep increase above 2 GPa. Above 4 GPa, $T_c^{onset}$ reaches a maximum value of 37K. The fact that the application of pressure dramatically enhances $T_c$ in FeSe indicates that the increase of $T_c$ should be related to the change in the local crystal structure, because the carrier density does not change largely with applied pressure. In fact, the dramatic enhancement of $T_c$ is successfully explained by investigating the structural parameter of "anion height" in the Fe layer [33,36,37].

As shown in Fig. 5, the pressure dependences of $T_c$ and the anion height for FeSe show an obvious correlation. This implies that $T_c$ of FeSe can be optimized by tuning the anion height. Interestingly, the anion height dependence of $T_c$ is applicable to all the Fe-based superconductors, not only for FeSe [36-38]. Fig. 6(a) is the anion height dependence of $T_c$ of the typical Fe-based superconductors [36]. A schematic image of anion height from the Fe layer is described in Fig. 6(b). The anion height dependence of $T_c$ shows a symmetric curve with a peak around 1.38 Å, as indicated by the hand-fitting curve. For FeSe, although the data point at ambient pressure does not agree with the unique curve, it corresponds above 2 GPa at which the anomaly was observed in the pressure dependence of $T_c$ as shown in Fig. 5. It indicates that intrinsic superconductivity in FeSe might be induced by the application of pressure above 2 GPa.

## 2-2. High-$T_c$ superconductivity in FeSe-related materials

As achieved by the application of the external pressure, the $T_c$ of FeSe can be enhanced by decreasing the anion height upon metal or molecule intercalations into the interlayer sites. Guo *et al*. reported that K-intercalated FeSe, $K_{0.8}Fe_2Se_2$ showed superconductivity with a $T_c$ above 30 K [13]. Interestingly, not only metal ions but also molecules such as $Li_x(NH_2)_y(NH_3)_{1-y}$ can be intercalated into the interlayer site of FeSe; $Li_x(NH_2)_y(NH_3)_{1-y}Fe_2Se_2$ shows high $T_c$ above 40 K [39]. To date, many FeSe-based superconductors with $T_c$ above 30 K have been reported [13, 39-46]. These facts indicate that the mechanisms of high-$T_c$ superconductivity in Fe-based compounds are essentially common in FeAs-based and Fe-chalcogenide-based superconductors.

## 2-3. FeTe

The crystal structure of FeTe is very analogous to that of FeSe. However, FeTe exhibits a magnetic/structural transition at 70 K. In fact, the physical properties of FeTe is quite different from those of FeSe [47,48]. FeTe is not superconducting except in the special case of the tensile-stressed FeTe thin film [49], which showed superconductivity at 13 K. Neutron-scattering studies indicate that the spin structure of FeTe is different from that of the Fe-pnictide parent compounds as shown Fig. 7 [47]. For FeSe and Fe-pnictide superconductors, the antiferromagnetic spin fluctuations with a wave vector $Q$s = (0.5, 0.5) was found to correlates with superconductivity [50-54]. On the other hand, FeTe shows magnetic wave vector $Q$d = (0.5, 0) [50,51]. Several reports indicated that the wave vector $Q$d is not favorable for superconductivity [55-57]. The emergence of a Fermi surface nesting associated with $Q$d could be induced by excess Fe at the interlayer site as described in Fig. 2(c). Excess Fe supplies a substantial amount of electrons and it has a magnetic moment. FeTe contains 7 – 25 % of excess Fe in its crystal structure, and thus the physical properties of FeTe depend on the content of excess Fe [55, 57].

2-4. FeTe-based superconductors

 2-4-1. FeTe$_{1-x}$Se$_x$

As mentioned above, FeTe exhibits an antiferromagnetic ordering associated with a lattice distortion at 70 K. Furthermore, FeTe possesses excess Fe (7 ~ 25 %) at the interlayer sites. However, partial Se substitutions suppress the low temperature structural/magnetic phase transition, and thereby produce superconductivity [11, 58, 59].

The effect of Se substitution other than the suppression of magnetic ordering is a reduction of excess Fe concentration. Due to both effects, bulk superconducting state can be realized. Fig. 8 shows the temperature dependence of magnetic susceptibility for FeTe$_{1-x}$Se$_x$ (a) around $T_N$, (b) around $T_c$. The long-range magnetic ordering was suppressed with increasing Se concentration and completely disappeared at $x = 0.15$. The $T_c$ is also gradually enhanced with increasing Se content and the optimal superconducting properties are obtained at a composition of FeTe$_{0.5}$Se$_{0.5}$ with $T_c$ of 14 K, which is the highest $T_c$ at ambient pressure among the FeTe-based superconductors. Furthermore FeTe$_{1-x}$Se$_x$ shows a positive pressure effect on $T_c$ as in the case of FeSe [60,61]. As depict in Fig. 9, a dome-shaped $T_c$ versus pressure curve arises and maximum $T_c$ of 23 K appears at 2 GPa [60].

2-4-2. FeTe$_{1-x}$S$_x$

As the partial Se substitution for Te induces superconductivity in FeTe, the S substitution for Te also suppresses the magnetism and induces superconductivity [12]. However, in the case of FeTe$_{1-x}$S$_x$, the optimal substitution cannot be achieved due to the low S/Te solubility limit as shown in Fig. 10 [62]. Although the antiferromagnetic ordering can be completely suppressed by the 20 % of S substitution, bulk superconductivity is not observed; only weak superconductivity is observed. This can be understood that the appearance of bulk superconductivity in FeTe$_{1-x}$S$_x$ is influenced by the higher content of excess Fe resulting from low S concentration. In particular, the FeTe$_{0.8}$S$_{0.2}$ samples synthesized using a solid state reaction method show a broad transition in temperature dependence of resistivity, and diamagnetic signal corresponding to superconductivity is not observed. In order to enhance the weak superconductivity, ingenious attempts have been carried out and exotic annealing effects were discovered.

2-5. Annealing effects of FeTe-based superconductors

Fig. 11(a) displays temperature dependence of resistivity for the FeTe$_{0.8}$S$_{0.2}$ sample kept in the air for several days. Although the as-grown sample does not show zero resistivity, the sample which was exposed to the air for several days shows zero resistivity with a sharp superconducting transition [63, 64]. After 2 years, the $T_c^{zero}$ reached 7.8 K. The superconducting signal of magnetic susceptibility also dramatically enhanced with increasing air-exposure time as shown Fig. 11(b).

Furthermore it is found that the weak superconductivity could be effectively improved by annealing in oxygen [65,66]. Fig. 12 shows the temperature dependence of magnetic susceptibility for the samples annealed with various annealing condition. Only the oxygen-annealed sample shows a large superconducting signal. Y. Kawasaki *et al*. reported that oxygen annealing is applicable to not only FeTe$_{1-x}$S$_x$ but also FeTe$_{1-x}$Se$_x$ [67]. Fig. 13 shows temperature dependence of magnetic susceptibility for O$_2$-annealed samples with various Se concentrations (a) around $T_N$, (b) around $T_c$. Only a 10% substitution of Te by Se completely suppressed the magnetic ordering and induced bulk superconductivity. The difference between as-grown and O$_2$-annealed samples is remarkable with the data in Fig. 8.

The phase diagrams based on the magnetic susceptibility measurement for the as-grown and O$_2$-annealed samples are shown in Fig. 14(a) and (b). The as-grown samples showed the long-range AFM in the range of $x \leq 0.15$ and weak superconductivity in $0.1 \leq x \leq 0.4$ where "weak superconductivity" means non-bulk (filamentary or partial) superconducting state interfered by excess Fe. Only the FeTe$_{0.5}$Se$_{0.5}$ sample was found to be a bulk superconductor. For O$_2$-annealed samples, the coexistence of AFM ordering and weak superconductivity was observed only for $x \leq 0.1$. As the long-range AFM ordering was completely suppressed, the O$_2$-annealed samples with $x \geq 0.1$ became bulk superconductors. It is clear from the Fig. 14 that the bulk superconducting region dramatically spreads via O$_2$-annealing.

Next, we discuss the origin of the appearance of bulk superconductivity on the basis of antiferromagnetic fluctuations. As mentioned in section 2.1, Antiferromagnetic spin fluctuations with the wave $Q$s = (0.5, 0.5) were observed in FeSe as well as the other FeAs. In contrast, FeTe exhibits antiferromagnetic wave vector $Q_d$ = (0.5, 0). From the neutron scattering measurements, it is revealed that the $Q$s and $Q_d$ are observed over a wide composition range where FeTe$_{1-x}$Se$_x$ exhibits weak superconductivity [56, 68]. For as-grown FeTe$_{1-x}$Se$_x$ samples, weak superconductivity was observed in the range of $0.1 \leq x \leq 0.4$, suggesting that the nesting vectors Qs and $Q_d$ coexist. It is expected that the nesting vector $Q$s becomes dominant for FeTe$_{0.5}$Se$_{0.5}$ where bulk superconductivity sets in. On the other hand, for the O$_2$-annealed samples, the bulk superconducting region extends down to $x = 0.1$, which implies that the nesting vector $Q$d is strongly suppressed by O$_2$-annealing. By investigating the oxygen annealing effect, it was concluded that the oxygen ions intercalated between superconducting layers play a key role in the suppression of magnetic wave vector $Q_d$ due to the compensation of the electrons given by the excess Fe ion and the appearance of bulk superconductivity.

2-6. Softchemical treatment for deintercalation of excess Fe

It was found that the alcoholic beverages induced superconductivity in $FeTe_{0.8}S_{0.2}$ [69]. In this report, the $FeTe_{0.8}S_{0.2}$ samples were immersed in red wine, white wine, beer, Japanese sake (rice wine), shochu (distilled spirit), and whisky, and then heated at 70ºC for 24 h. The obtained shielding volume fractions are summarized in Fig. 15 as a function of ethanol concentration. The shielding volume fractions of the samples heated in water-ethanol mixtures are between 6 and 9 %. In contrast, the shielding volume fraction of the samples heated in alcoholic beverages is 21 - 63 %, significantly larger than that with the water-ethanol mixtures. In the first report, we concluded that alcoholic beverages, particularly red wine, were more effective in inducing superconductivity in $FeTe_{0.8}S_{0.2}$ than water, although the exact mechanism of how they acted to enhance the superconductivity in $FeTe_{1-x}S_x$ remained unsolved. This unique annealing effect was also confirmed by $FeTe_{1-x}Se_x$ sample. Fig. 16 indicates alcoholic annealing effect for $FeTe_{0.9}Se_{0.1}$. In the case of $FeTe_{0.9}Se_{0.1}$, the highest volume fraction was observed by red wine sample and the smallest value among the alcoholic beverages was obtained with shochu. This behavior is similar to the case of $FeTe_{1-x}S_x$.

Recently, we have successfully clarified that the mechanism of inducement of superconductivity in $FeTe_{1-x}S_x$ by alcoholic beverages is the deintercalation of excess Fe from the interlayer sites [70]. Utilizing a technology of metabolomic analysis, we performed the investigation of the ingredient in the alcoholic beverage and found that the solutions of malic acid, citric acid, and $\beta$-alanine also induced the superconductivity in $FeTe_{0.8}S_{0.2}$ as depicted in Fig. 17(a) and (b). Additionally, inductively-coupled plasma spectroscopy analysis indicated that Fe ion was deintercalated from the sample to these solutions as shown Fig. 17(c). These results suggest that part of the excess Fe was deintercalated from the interlayer sites. Therefore, alcoholic beverage annealing suppresses the magnetic moment of excess Fe by deintercalating the excess Fe and, hence superconductivity is achieved. The similar situations, enhancement of the superconductivity by the deintercalation of excess Fe, were reported [71, 72] Thus the technique can be generally applied to the layered Fe-chalcogenide superconductors.

## 3. BiS$_2$-based layered superconductors

Discovery of a common superconducting layer is very important because many analogous superconductors can be designed by changing the spacer layers as have the high-$T_c$ cuprates and the Fe-based superconductors. Recently, we found that the $BiS_2$ layer can be a basic superconducting layer of new class of the layered superconducting family.

The first $BiS_2$-based superconductor is $Bi_4O_4S_3$ [20]. The determined crystal structure is shown in Fig. 18(a). The crystal structure is composed of a stacking of rock-salt-type

$BiS_2$ layers and $Bi_4O_4(SO_4)_{1-x}$ layers (blocks) where $x$ indicates the existence of the

defects of $SO_4^{2-}$ ions at the interlayer sites. The parent phase ($x = 0$) is $Bi_6O_8S_5$, and

$Bi_4O_4S_3$ is expected to have about 50 % defects of the $SO_4^{2-}$ site (x = 0.5). In layered

materials, such defects of the interlayer sites are often observed.

Temperature dependence of resistivity is displayed in Fig. 19. A gradual

decrease of resistivity was observed below 8.6 K and the zero-resistivity state was

observed at 4.5 K. Magnetic susceptibility measurements indicate that $Bi_4O_4S_3$ was a

bulk superconductor. To investigate the origin of superconductivity in $Bi_4O4S_3$, we

performed band calculations. The band calculation indicated that $Bi_4O_4S_3$ ($x = 0.5$) was

metallic while the parent phase of $Bi_6O_8S_5$ ($x = 0$) was found to be a band insulator with

$Bi^{3+}$. For $Bi_4O_4S_3$, the Fermi level lies within the bands which mainly originate from the

Bi 6p orbitals as displayed in Fig. 20. In particular, the Fermi level is just on the peak position of the partial density of states of the Bi 6p orbital within the $BiS_2$ layer.

The second $BiS_2$-based system is $ReO_{1-x}F_xBiS_2$ (Re = Rare earth). So far,

superconductivity is observed in $LaO_{1-x}F_xBiS_2$ [24], $NdO_{1-x}F_xBiS_2$ [25], $CeO_{1-x}F_xBiS_2$

[73] and $PrO_{1-x}F_xBiS_2$ [74]. The crystal structure of $LaO_{1-x}F_xBiS_2$ is show in Fig. 18(b).

These materials also possess the $BiS_2$ layers as superconducting layers. The structure is simpler than that of $Bi_4O_4S_3$; hence, we can consider this system as a prototype of $BiS_2$-based superconductors. Temperature dependences of resistivity under magnetic fields for $LaO_{1-x}F_xBiS_2$ and $NdO_{1-x}F_xBiS_2$ are shown in Figs. 21(a) and (b), respectively. Super conducting transition occurs at 10.6 and 5.6 K for $LaO_{1-x}F_xBiS_2$ and $NdO_{1-x}F_xBiS_2$. We note that the blocking layer of $La_2O_2$ or $Nd_2O_2$ is analogous to those in FeAs-1111 system such as LaFeAsO. Owing to the analogy of crystal structure between the $BiS_2$-based and the Fe-based superconductors, we will easily design new $BiS_2$-based superconductors by changing the blocking layers. We strongly expect that further enhancements of $T_c$ are achieved by changing the blocking layers in the $BiS_2$-based superconductor and the discovery of new $BiS_2$-based materials open a new field in physics and chemistry of low-dimensional superconductors. In fact, some preliminary reports have indicated the possibility of unconventional superconductivity in the $BiS_2$-based superconductors [75-78].

4. Conclusion

In this review, we introduced the crystal structure and physical properties of remarkable layered chalcogenide superconductors. Chalcogenides tend to crystallize in a layered structure; hence, the intercalations/deintercalations of ions or molecules at the interlayer site dramatically changes the physical properties and induces exotic superconductivity. The most remarkable family is the Fe chalcogenides, which is the simplest Fe-based superconductor. In this series, the key factors to induce superconductivity are the suppression of antiferromagnetism of Fe planes and the reduction of magnetic moment of excess Fe at the interlayer site. The later, reduction of magnetic moment of excess Fe can be achieved by oxygen intercalation via annealing in oxygen condition or deintercalation of excess Fe via annealing in acid. Interestingly, red wine is the most effective than any other solution. At the end, we introduced the newly discovered $BiS_2$-based superconducting family. The $BiS_2$ layer is likely to play an important role of the superconductivity, as $CuO_2$ plane of cuprates and FeAn (FeAs, FeP, FeSe or FeTe) layers of Fe-based superconductors. We will be able to create new $BiS_2$-based superconductors with various blocking layers. We believe that unidentified exotic chalcogenide superconductors other than the families introduced here exist and are waiting to be discovered in near future.

Acknowledgements

The authors thank the group members of National Institute for Materials Science: Mr. Hara H, Mr. Demura S, Mr. Watanabe T, Dr. Fujioka M, Dr. Denholme S J, Dr.

Okazaki H, Dr. Ozaki T, Dr. Takeya H and Dr. Yamaguchi T for useful discussions and experimental help. The authors thank the group members of Tokyo Metropolitan University: Mr. Hamada K, Mr. Izawa H and Dr. Miura O for fruitful discussions and experimental supports. This work was partly supported by a Grant-in-Aid for Scientific Research (KAKENHI) and a Strategic International Collaborative Research Program (SICORP), Japan Science and Technology Agency.

Figure Captions

Fig. 1. Crystal structure of the typical layered chalgenids, (a) $TiSe_2$, (b) $BiSe_2$ and (c) FeSe.

Fig. 2. Crystal structure of the typical Fe-based superconductors (a) LaFeAO, (b) FeSe and (c) $Fe_{1+d}Te$. For $Fe_{1+d}Te$, excess Fe occupies the interlayer sites with an occupancy of 7 – 25 % as indicated with partially-filled circles.

Fig. 3. Temperature dependence of resistivity for FeSe. The insets show temperature dependence of resistivity under high magnetic fields and the obtained field-temperature phase diagram. Reprinted from ref. 10.

Fig. 4. Temperature dependence of resistivity for FeSe under high pressure measured using (a) a piston-cylinder cell and (b) an indenter cell.

Fig. 5. Pressure dependence of both the $T_c$ and the Se height for FeSe.

Fig. 6. (a) Anion height dependence of $T_c$ of the typical Fe-based superconductors. The $T_c$ is the highest value in the system. (b) Schematic image of anion height.

Fig. 7. Schematic in-plane spin structure of FeTe and $SrFe_2As_2$. Reprinted from ref. 47

Fig. 8. Temperature dependence of normalized magnetic susceptibility for $FeTe_{1-x}Se_x$ (a) around $T_N$ (b) around $T_c$

Fig.9. Pressure evolution of $T_c$ as a function of applied pressure. Reprinted from ref. 60.

Fig. 10. Nominal $x$ dependence of $x_E$ determined by EPMA. Reprinted from ref. 62.

Fig. 11. Temperature dependence of (a) resistivity and (b) magnetic susceptibility for $FeTe_{0.8}S_{0.2}$.

Fig.12. Temperature dependence of magnetic susceptibility for $FeTe_{0.8}S_{0.2}$ with several annealing conditions.

Fig. 13. Temperature dependence of magnetic susceptibility for $O_2$-annealed samples with various Se concentrations (a) around $T_N$, (b) around $T_c$.

Fig. 14. Phase diagrams showing $T_c$ and $T_N$ as a function of $x$ for (a) as-grown $FeTe_{1-x}Se_x$ and (b) $O_2$-annealed $FeTe_{1-x}Se_x$.sdsd

Fig. 15. The shielding volume fraction of $FeTe_{0.8}S_{0.2}$ samples heated in various liquids as a function of ethanol concentration.

Fig. 16. (a) Temperature dependence of magnetic susceptibility for $FeTe_{0.9}Se_{0.1}$ with various annealing conditions. (b) The shielding volume fraction of $FeTe_{0.1}Se_{0.1}$ samples heated in various liquids as a function of pH.

Fig. 17. (a) Temperature dependence of magnetic susceptibility for $FeTe_{0.8}S_{0.2}$ annealed in various solutions. (b) The shielding volume fractions for the samples annealed in various liquids as a function of pH. (c) The pH dependence of Fe concentration dissolved in solutions after annealing with the sample.

Fig. 18. (a) Crystal structure of $Bi_4O_4(SO_4)_{1-x}Bi_2S_4$. The composition of $x = 0.5$ corresponds to $Bi_4O_4S_3$. (b) Crystal structure of $LaOBiS_2$.

Fig. 19. Temperature dependence of resistivity for $Bi_4O_4S_3$ under magnetic fields up to 5 T.

Fig. 20. The band structure for $Bi_4O_4S_3$. The radius of the circles represent the weight of the Bi 6p orbitals within the $BiS_2$ layer. The blue and red hatches indicate the bands having mainly $p_x$ and $p_y$ characters, respectively. Right panel: the partial density of states of the Bi 6p orbital within the $BiS_2$ layer.

Fig. 21. (a) Temperature dependence of resistivity for $LaO_{0.5}F_{0.5}BiS_2$ under magnetic fields up to 5 T. (b) Temperature dependence of resistivity for $NdO_{0.7}F_{0.3}BiS_2$ under magnetic fields up to 4 T.

Fig. 1

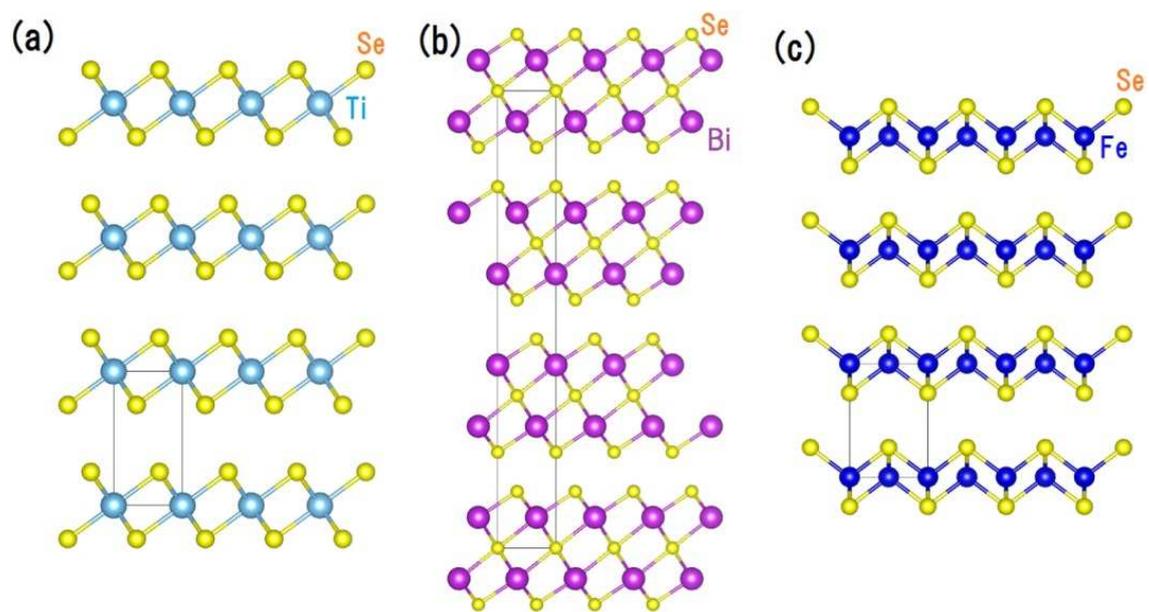

Fig 2

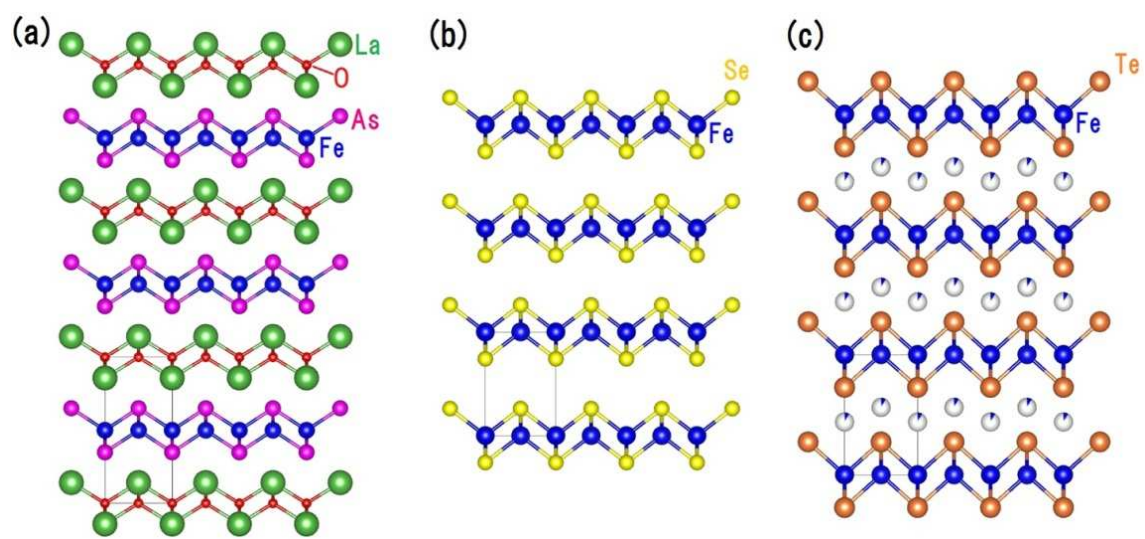

Fig. 3

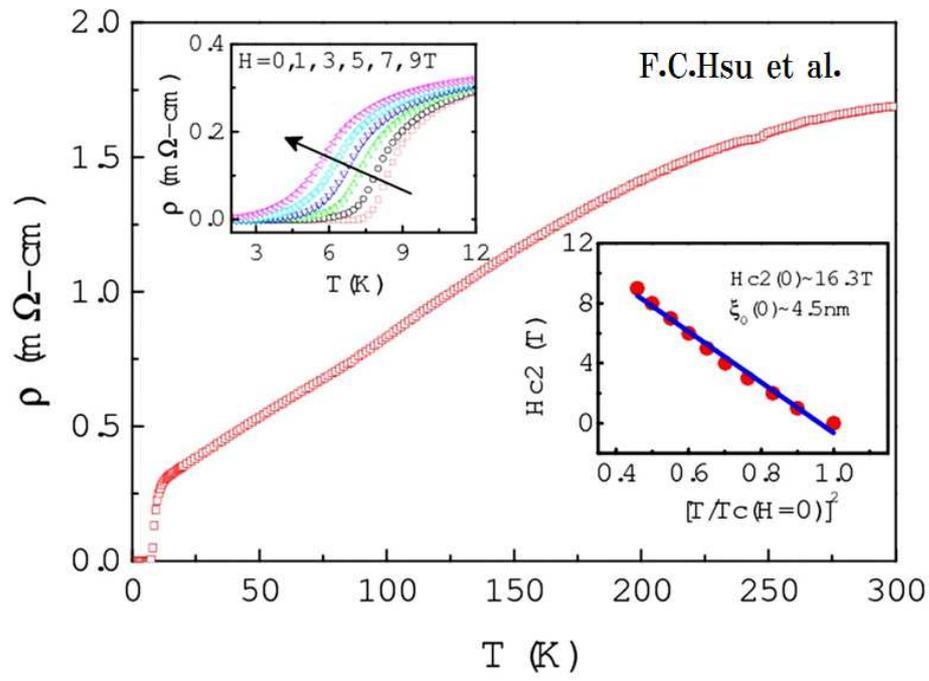

Fig. 4

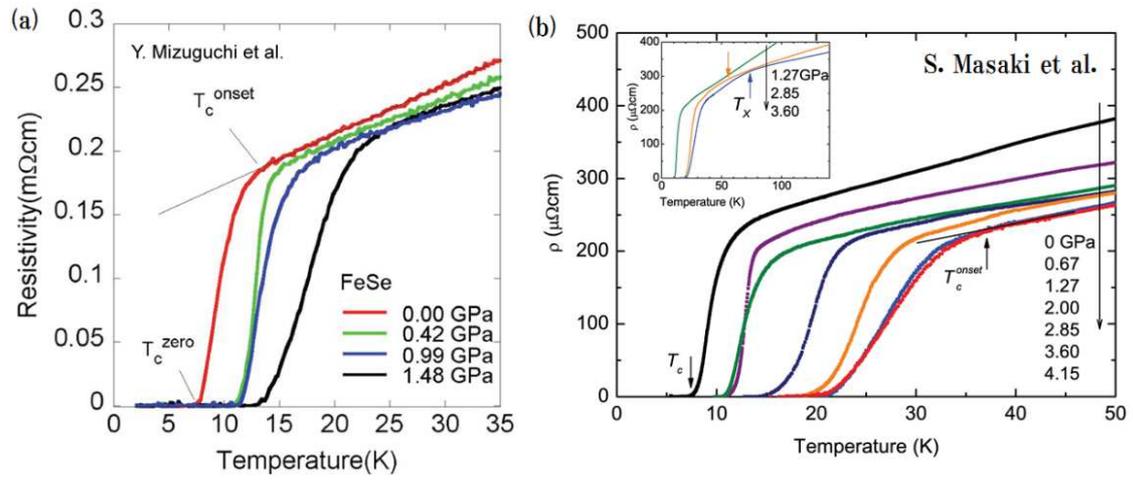

(a) Y. Mizuguchi et al.

$T_c^{onset}$

$T_c^{zero}$

FeSe

| | |
|---|---|
| — | 0.00 GPa |
| — | 0.42 GPa |
| — | 0.99 GPa |
| — | 1.48 GPa |

(b) S. Masaki et al.

$T_x$

1.27GPa
2.85
3.60

$T_c^{onset}$

$T_c$

0 GPa
0.67
1.27
2.00
2.85
3.60
4.15

Fig. 5

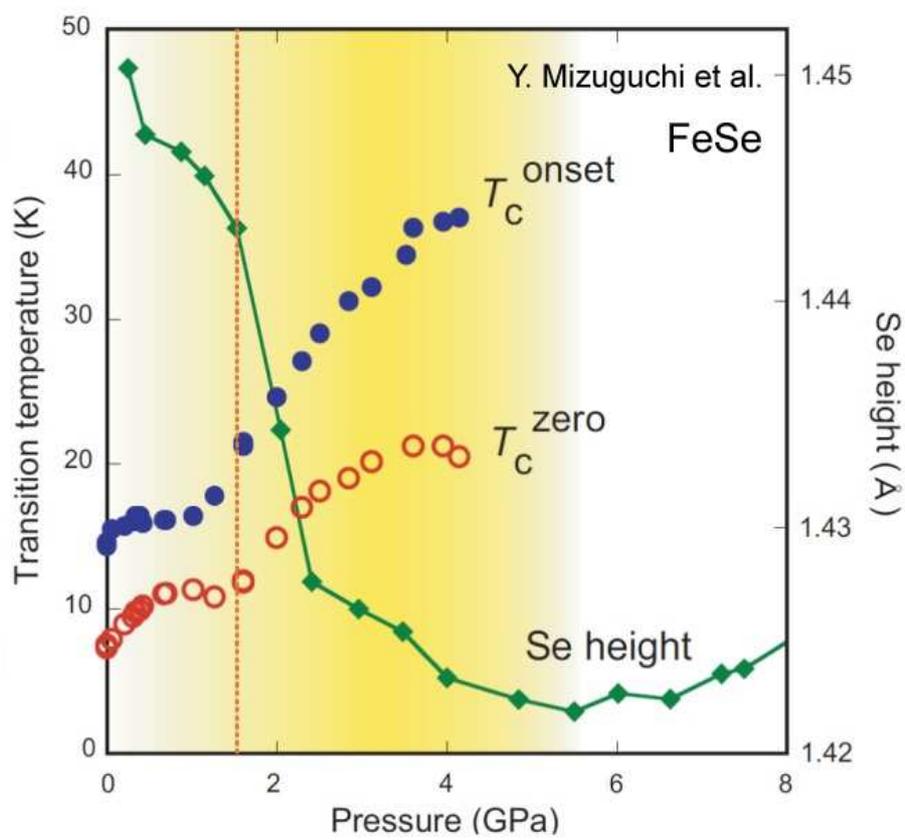

Fig. 6

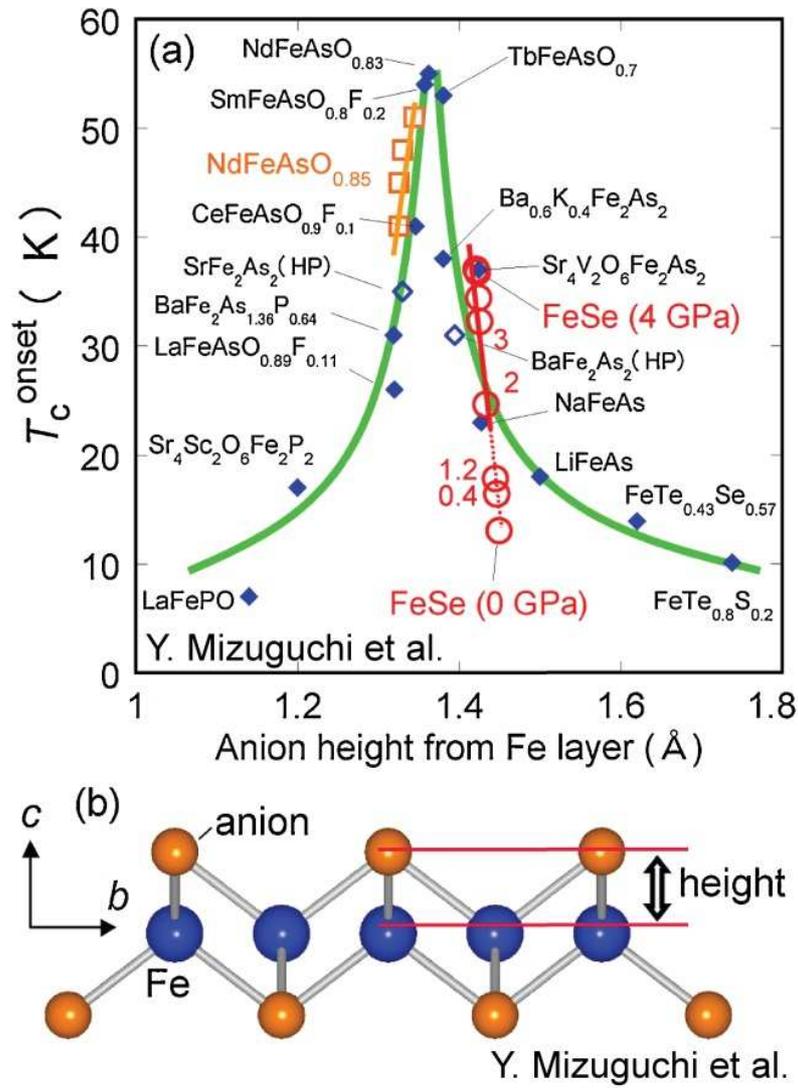

Fig. 7

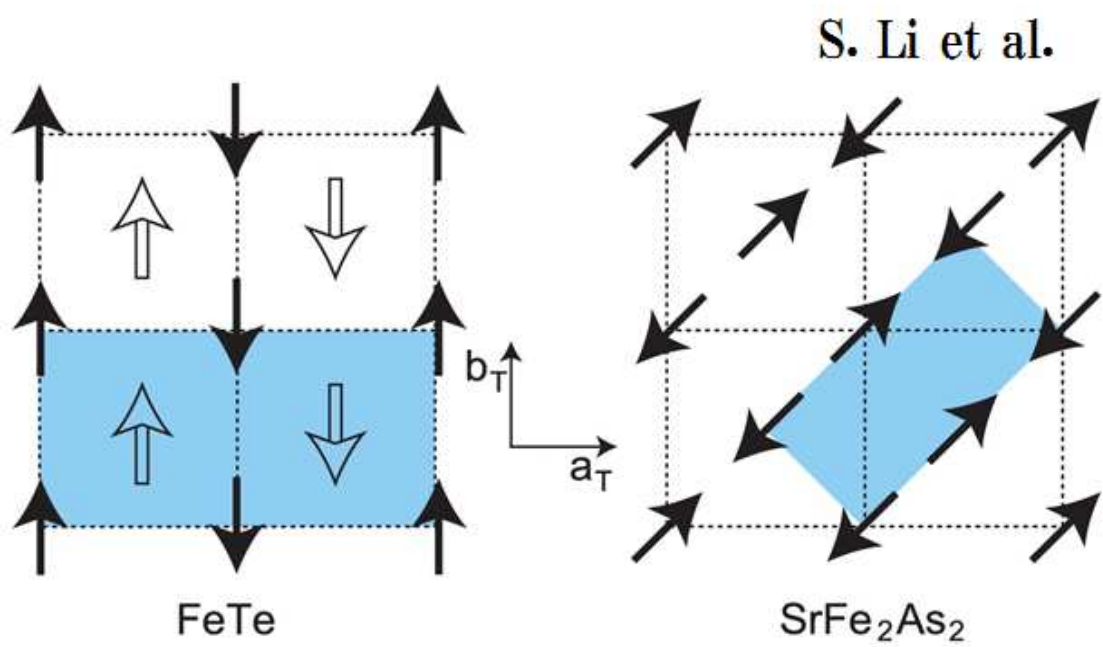

S. Li et al.

FeTe          SrFe$_2$As$_2$

Fig. 8

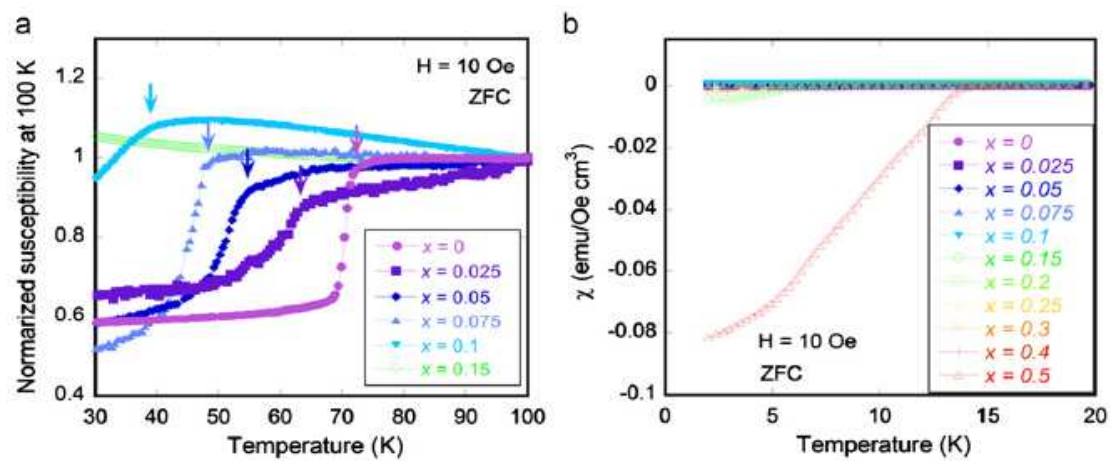

Fig. 9

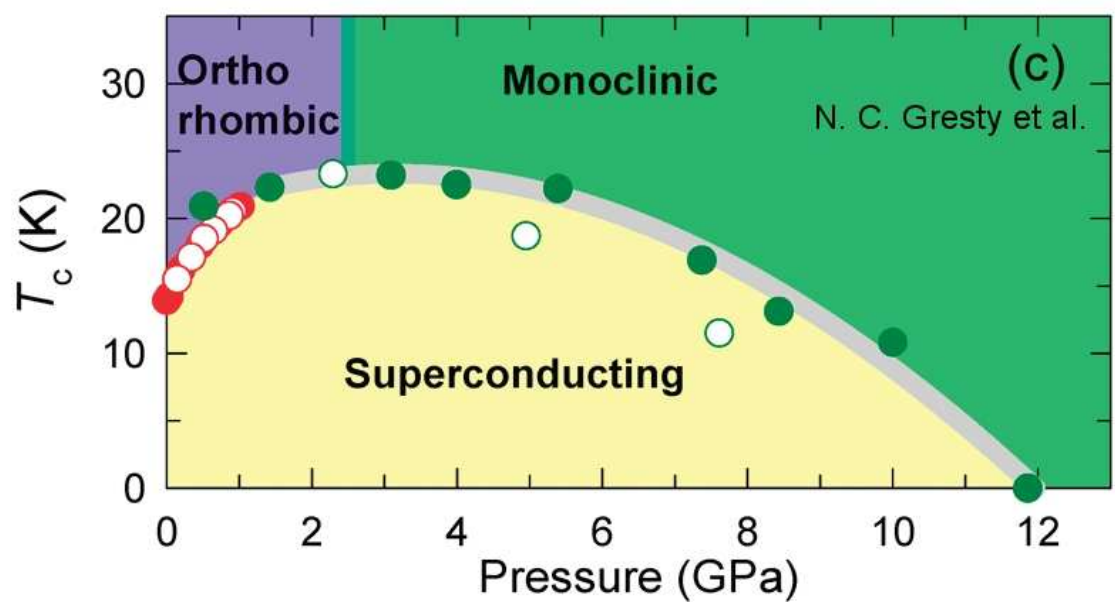

Fig. 10

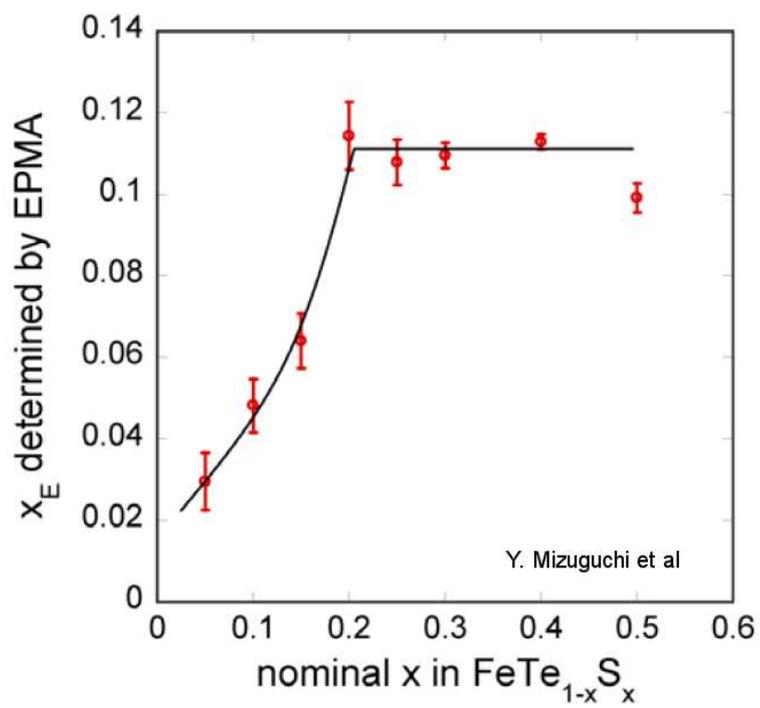

Fig. 11

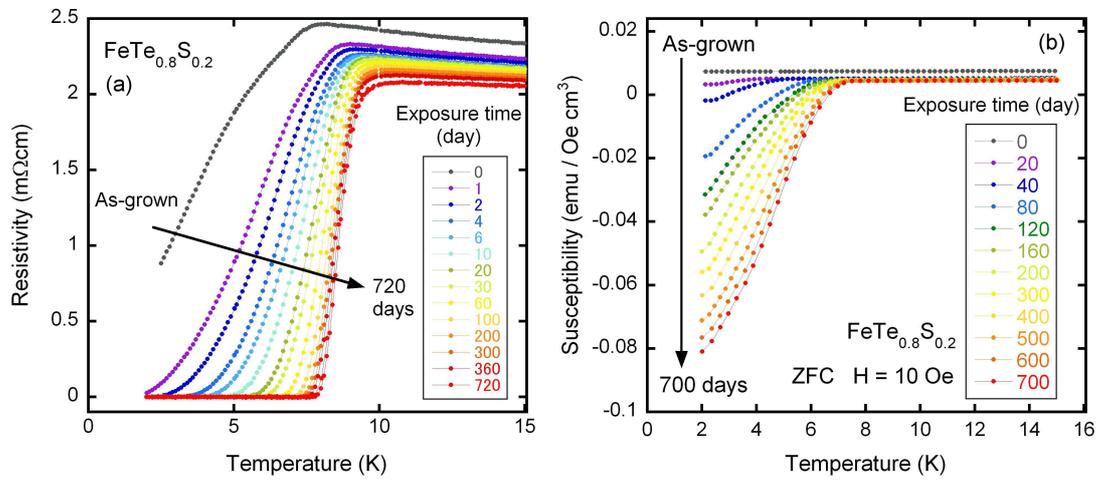

Fig. 12

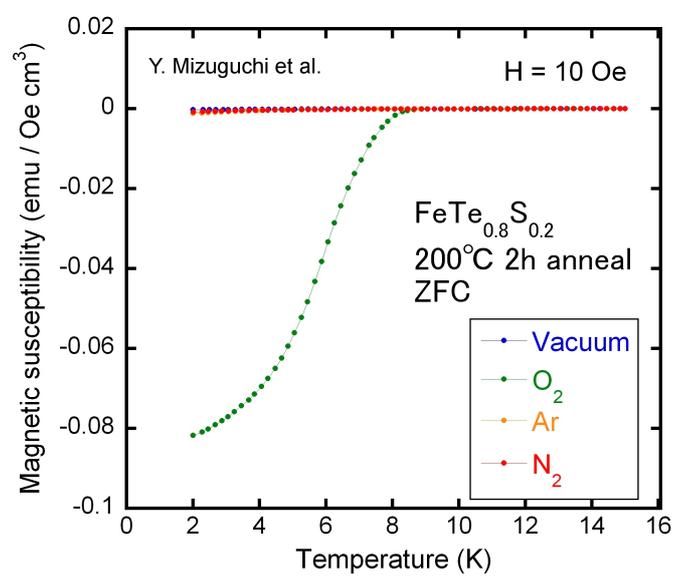

Fig. 13

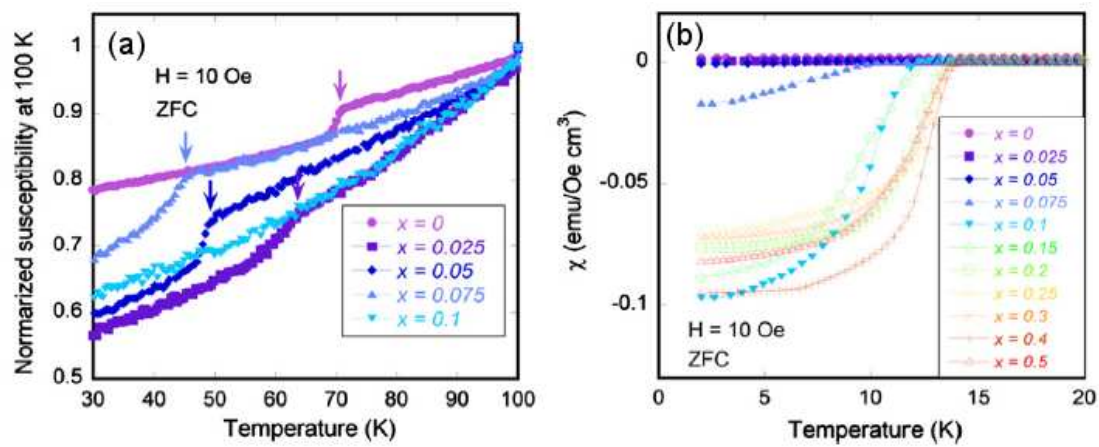

Fig. 14

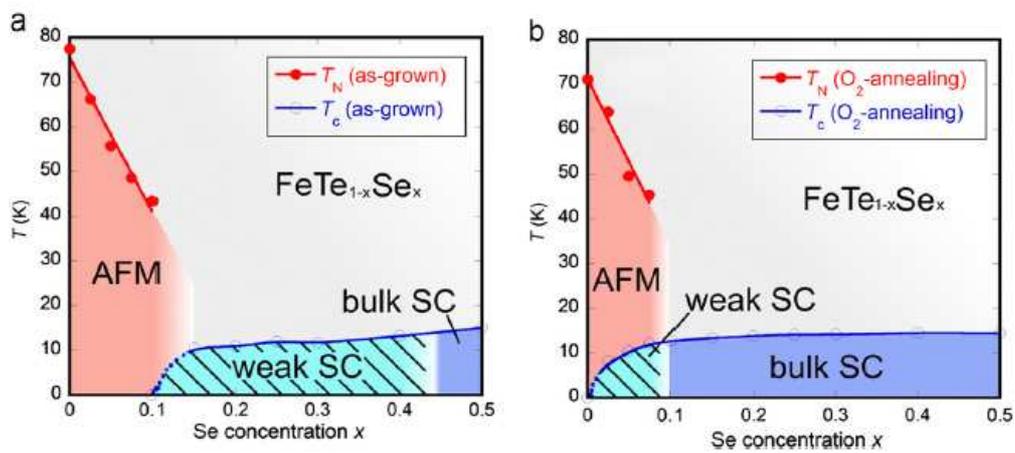

Fig. 15

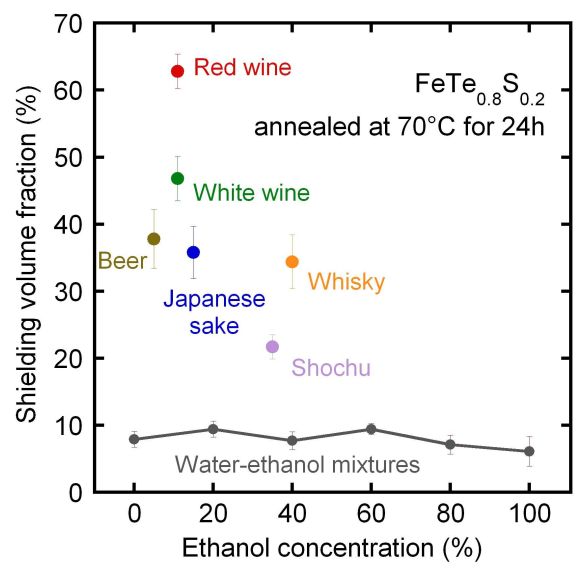

Fig. 16

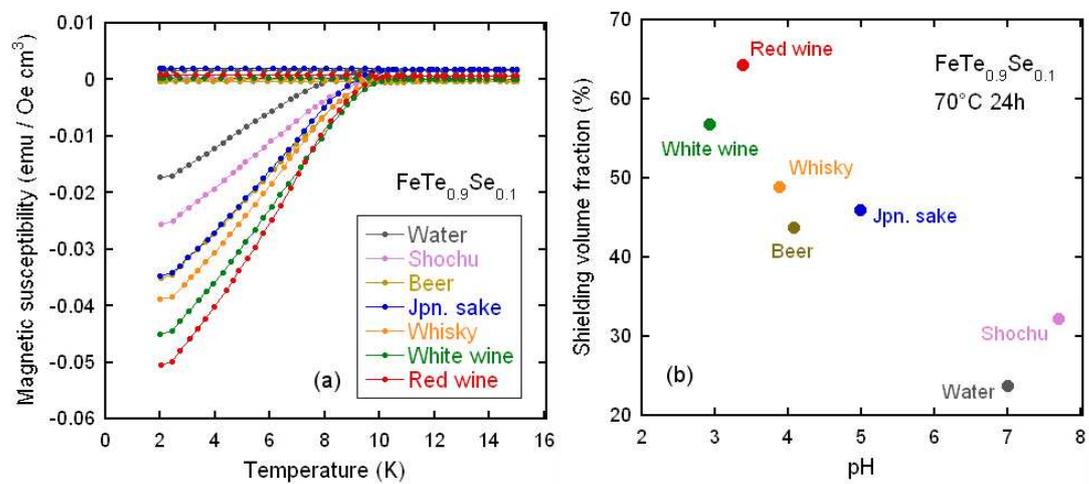

Fig. 17

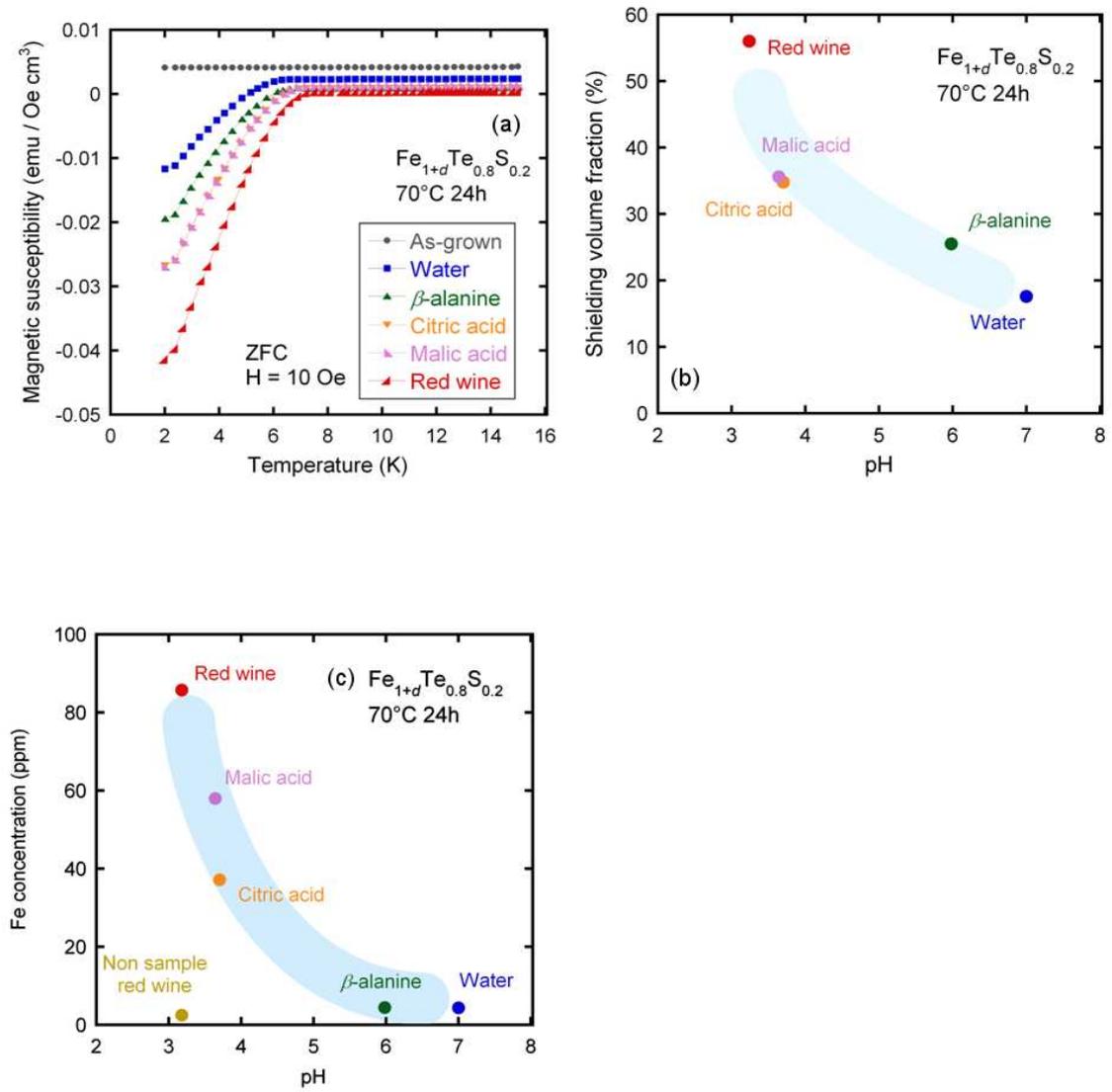

Fig. 18

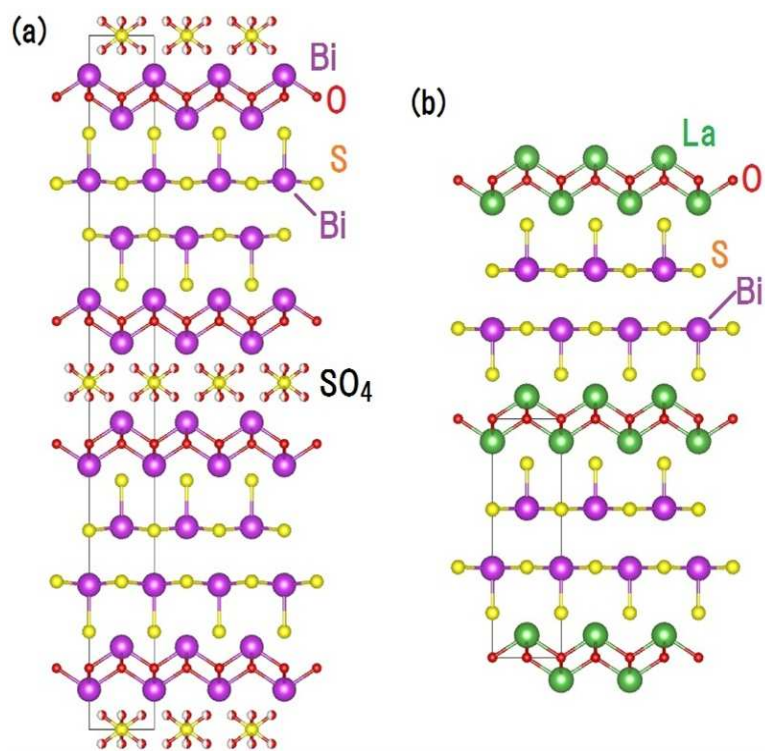

Fig. 19

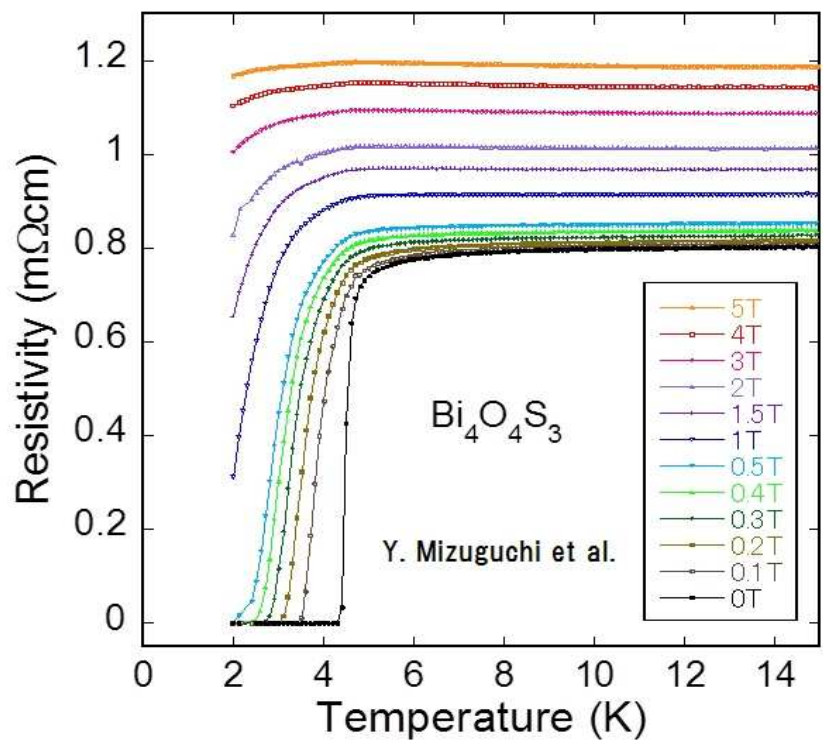

Fig. 20

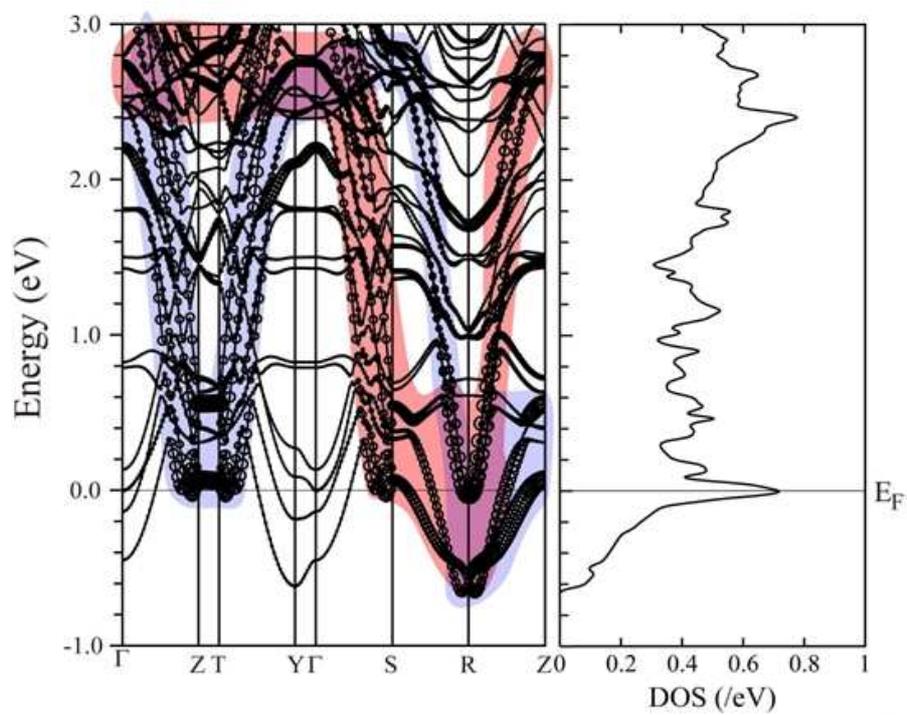

Y. Mizuguchi et al.

Fig. 21

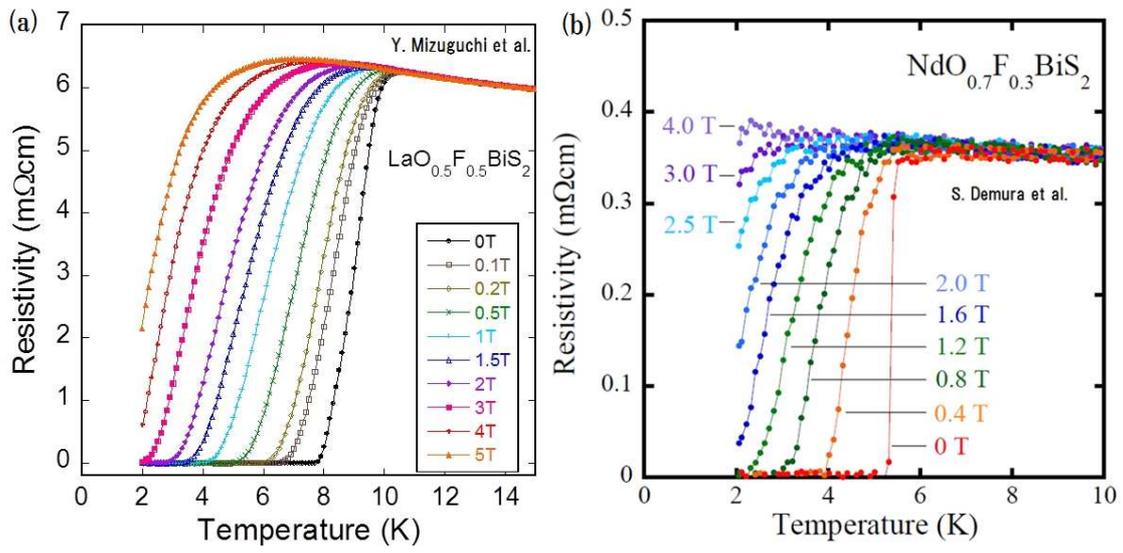